\documentclass[12pt]{article}

\usepackage{newtxtext,newtxmath}
\usepackage[utf8x]{inputenc}
\usepackage[english]{babel}
\usepackage{amsmath,amsbsy,amsfonts}
\usepackage{graphicx}
\usepackage{subfig}
\usepackage{color}
\usepackage{float}
\usepackage{cite}
\usepackage[left=1cm,right=1cm,top=2cm,bottom=2cm]{geometry}
\usepackage{physics}
\usepackage[colorlinks=true, allcolors=red]{hyperref}

\numberwithin{equation}{section}

\begin{document}

\title{Airy eigenstates and their relation to coordinate eigenstates}
\author{Jorge A. Anaya-Contreras$^1$, Arturo Z\'u\~niga-Segundo$^1$ and H\'ector M. Moya-Cessa$^2$\\
$^1$\small Instituto Polit\'ecnico Nacional, ESFM, Departamento de F\'isica. Edificio 9, Unidad Profesional ''Adolfo L\'opez Mateos"\\ CP 07738 CDMX, Mexico\\
$^2$\small {Instituto Nacional de Astrof\'isica, \'Optica y Electr\'onica, 72840 Sta. Mar\'ia Tonantzintla, Puebla, Mexico}}

\date{\today}

\maketitle

\begin{abstract}
We study the eigenvalue problem for a linear potential Hamiltonian and, by writing  Airy equation in terms of momentum and position operators  define Airy states. We give a solution of the Schr\"odinger equation for the symmetrical linear potential in terms of the squeeze and displacement operators. Finally, we write the unit operator in terms of Airy states and find a relation between them and position and momentum eigenstates.
\end{abstract}

\section{Introduction}

The study of the Airy functions has attracted a lot attention in many branches of physics and applied mathematics. This because  Airy functions are eigenfunctions of a Hamiltonian with a linear potential \cite{sakurai,gea_1999}, as well as for its non-spreading and bending properties \cite{ZHUKOVSKY20117966} in free space as well as in inhomogeneous, {\it i.e.,} time-varying linear potential \cite{berry_1979}. Airy beams have been introduced in optics due to their intriguing properties:   self-healing \cite{Broky:08} and self-accelerating along a parabolic trajectory \cite{PhysRevLett.99.213901,Siviloglou:07,ChavezCerda2011}. Taking into account these unique characteristics, the Airy beams and Airy functions have various applications in many areas, such as light-sheet microscopy \cite{Vettenburg2014}, electron Airy waves \cite{Voloch-Bloch2013} and Stark effect \cite{Avron1977,Matin2014}.\\

The coordinate representation of the Schr\"odinger equation is the more usual way to board and to set problems in quantum mechanics. For instance the eigenfunctions of the harmonic oscillator, are represented by the braket $\langle x\vert n\rangle$, where the kets $\vert n\rangle$ with $n=0,1,2,\cdots$, are called number or Fock states (see for instance \cite{Leonhardt}). And any ket $\vert g\rangle$ can be expanded in terms of these number states. The basis set of kets $\vert n\rangle$ is discrete, however, there are also continuous bases. The questions we want to answer here are, Is it possible to have an Airy basis similar to number states? and Is it possible to relate Airy eigenstates to position eigenstates?

In general, those are complex questions. The main aim of this paper is to try to answer both. We present an explicit eigenstate of the linear symmetrical linear potential Hamiltonian. These states are orthogonal and their eigenvalues are not equally spaced. We present two numerical examples in order to illustrate our findings and the relation between the Airy basis with the usual coordinate eigenstates. Finally, we present a relation that allows to obtain the Airy states from the application of an exponential operator to the vacuum state.

\section{Airy states}

Airy functions $\hbox{Ai}(x)$ may be defined by the following Fourier transform,
\begin{equation} \label{eq001}
	\hbox{Ai}(x)=\frac{1}{2\pi}\int_{-\infty}^{+\infty}dt\;\hbox{e}^{i\left(\frac{t^{3}}{3}+xt\right)}\;.
\end{equation}
It is well known that Airy functions cannot be normalized, however they have the interesting property
\begin{equation} \label{eq003}
	\int_{-\infty}^{+\infty}dx\;\hbox{Ai}(y-x)\hbox{Ai}(y'-x)=\delta(y-y')\;,
\end{equation}
where $\delta(y-y')$ is the Dirac delta function.\\

In order to obtain the Airy state, we consider the Schr\"odigner equation for a single particle under a linear potential $V(x)=\vert k\vert x$, and $-\infty<x<\infty$, (for the sake of simplicity we set $m=\hbar=1$), that is 
\begin{equation} \label{eq004}
	-\frac{1}{2}\frac{d^{2}}{dx^{2}}\psi(x)+ \vert k\vert x\psi(x)=E\psi(x)\;.
\end{equation}
By the substitution, $z=\sqrt[3]{2\vert k\vert}\left(x-E/\vert k\vert\right)$, the above equation is transformed into the Airy equation
\begin{equation} \label{eq005}
	\frac{d^{2} \phi(z)}{dz^{2}}-z\phi(z)=0\;,
\end{equation}
whose solution is determined by an Airy function, $\hbox{Ai}(z)$ or $\hbox{Bi}(z)$. In this situation the solution $\hbox{Bi}(z)$ is not acceptable, since $\hbox{Bi}(z)$ goes to infinity as $z$ grows. Then the solution of equation (\ref{eq004}) is given by
\begin{equation} \label{eq006}
	\psi_{{}_E}(x)=\sqrt[6]{2\vert k\vert}\hbox{Ai} \left(\sqrt[3]{2\vert k\vert}\left(x-\frac{E}{\vert k\vert}\right)\right)\;,
\end{equation}
where we have added the constant $\sqrt[6]{2\vert k\vert}$ in order to the equation (\ref{eq006}) in order to satisfy  relation (\ref{eq003}).

Although the spectral properties of operators related to the linear potential (Stark effect), was proposed in reference \cite{Avron1977}, we prefer to use the squeeze $\hat{S}(r)$ and displacement $\hat{D}(\alpha)$ operators, with parameters $r= \ln\sqrt[3]{2\vert k\vert}$ and $\alpha =E/({\sqrt{2}\vert k\vert})$, respectively. Where the action of the above operators on arbitrary function is well known \cite{Leonhardt}. We may rewrite the solution of the Schr\"odinger equation (\ref{eq006}) as 
\begin{equation} \label{eq007}
	\psi_{_{E}}(x)=\hat{D}\left(\frac{E}{\sqrt{2}\vert k\vert}\right)\hat{S} \left(\ln\sqrt[3]{2\vert k\vert}\right) \hbox{Ai}(x)\;.
\end{equation}
If we set $\psi_{_{E}}(x)=\langle x\vert \psi_{_{E}}\rangle$, we have
\begin{equation} \label{eq008}
	\vert\psi_{_{E}} \rangle=\hat{D}\left(\alpha\right)\hat{S} \left(r\right)\vert\hbox{Ai} \rangle\;,
\end{equation}
where $\vert\hbox{Ai}\rangle$ is the Airy state. We wish to express the Schr\"odinger equation (\ref{eq004}) in terms of the ket $\vert\hbox{Ai}\rangle$, in order to establish an eigenvalue equation. Inserting equation (\ref{eq008}) into equation (\ref{eq004}) we obtain
\begin{equation} \label{eq011} 
	\hat{S}^{\dagger}\left(r\right)\hat{D}^{\dagger}\left(\alpha\right)\left(\frac{\hat{p}^{2}}{2}+\vert k\vert\hat{x}\right)\hat{D}\left(\alpha\right)\hat{S} \left(r\right)\vert\hbox{Ai}\rangle=E\vert\hbox{Ai}\rangle\;,
\end{equation}
where we rewrite the Hamiltonian (\ref{eq004}) in terms of the coordinate and momentum operators $\hat x$ and $\hat{p}$, respectively. It may be shown that \cite{Leonhardt}
\begin{eqnarray} \label{eq0012}
	\hat{D}^{\dagger}\left(\alpha\right)\left(\frac{\hat{p}^{2}}{2}+\vert k\vert\hat{x}\right)\hat{D}\left(\alpha\right)&=&\frac{\hat{p}^{2}}{2}+\vert k\vert\hat{x}+E\;,\nonumber\\
	\hat{S}^{\dagger}\left(r\right)\left(\frac{\hat{p}^{2}}{2}+\vert k\vert\hat{x}\right)\hat{S}\left(r\right)&=&\frac{(2\vert k\vert)^{\frac{2}{3}}}{2}\left(\hat{p}^{2}+\hat{x}\right)\;,
\end{eqnarray}
where we have
\begin{equation} \label{eq0013}
	(\hat{p}^{2}+\hat{x})\vert\hbox{Ai}\rangle=0\;,
\end{equation}
that confirms that the ket $\vert\hbox{Ai}\rangle$ satisfies the Airy equation. But does the Airy equation represents an eigenvalue problem? This question may be answered by using the displaced ket $\vert\gamma,\hbox{Ai}\rangle=\hbox{e}^{-i\gamma \hat{p}}\vert\hbox{Ai}\rangle$, and it is easy to see that
\begin{equation} \label{eq0014}
	\left(\hat{p}^{2}+\hat{x}\right)\vert\gamma,\hbox{Ai}\rangle=\gamma\vert\gamma,\hbox{Ai}\rangle\;,
\end{equation}
we have thus shown that any displaced Airy ket $\vert\gamma,\hbox{Ai}\rangle$ are eigenvectors of the Airy equation with eigenvalue $\gamma$, in particular when $\gamma=0$ as in equation (\ref{eq0013}). If we assume that
\begin{equation} \label{eq0015}
	\left(\frac{\hat{p}^{2}}{2}+\vert k\vert\hat{x}\right)\vert\psi_{_{E'}} \rangle=E'\vert\psi_{_{E'}} \rangle\;,
\end{equation}
where $\vert\psi_{_{E'}} \rangle=\hat{D}\left(\alpha\right)\hat{S} \left(r\right)\vert\gamma,\hbox{Ai}\rangle$, we have
\begin{equation} \label{eq0016}
	E'=\frac{(2\vert k\vert)^{\frac{2}{3}}}{2}\gamma+E\;,
\end{equation}
and, by making $\gamma=0$, we recover equation $\ref{eq011}$.

On the other hand, we have a completeness  relation
\begin{equation}
\label{eq0019}
\int_{-\infty}^{+\infty} d\gamma\vert\gamma,\hbox{Ai}\rangle\langle \gamma,\hbox{Ai}\vert=\hat{I}\;,
\end{equation}
that  may be verified by using $\langle x\vert\hat{I}\vert x'\rangle=\delta(x-x')$, and equation (\ref{eq003}).\\

To finish this section, let us consider again the Schr\"odigner equation for a single particle under a linear potential $V(x)=-\vert k\vert x$, for $-\infty<x<\infty$, whose Hamiltonian may be reduced to the last one by considering
\begin{equation} \label{eq0020}
	(-1)^{\hat{n}} \left( \frac{\hat{p}^{2}}{2} -\vert k\vert\hat{x} \right)(-1)^{\hat{n}}=\frac{\hat{p}^{2}}{2}+\vert k\vert\hat{x}\;,
\end{equation}
where $(-1)^{\hat{n}}$ is the so-called parity operator. Then, we have
\begin{equation} \label{eq0021}
	\vert\psi_{_{E}}\rangle=(-1)^{\hat{n}}\hat{D}\left(\frac{E}{\sqrt{2}\vert k\vert}\right)\hat{S}\left(\ln\sqrt[3]{2\vert k\vert}\right)\vert\hbox{Ai}\rangle\;.
\end{equation}
Therefore, we may conclude that the algebraic Airy problem does not depend on the sign of $k$.

\section{Airy basis state} 

The fact that the optical paraxial Helmholtz equation is mathematically equivalent to a Schr\"odinger equation, may be responsible for the many analogies that are found between quantum physics and classical optics. We will now consider the collapses and revivals occurring for Airy-beam propagation
in a symmetrical linear potential and waveguide respectively, using an Airy bases states.\\ 

Let us consider a symmetrical linear potential $V=\lambda\vert x\vert$, where $\lambda>0$, the Schr\"odigner equation for a single particle under  this potential reads,
\begin{equation}
	\label{eq3.1}
	-\frac{1}{2}\frac{d^{2}}{dx^{2}}\psi(x)+\lambda\vert x\vert\psi(x)=E\psi(x)\;,
\end{equation}
again, we have used appropriately scaled units $m=\hbar=1$ to simplify the equation. Now we are going to solve for all values of $x$.\\

\noindent Let us assume $x\geq 0$, the above Schr\"odinger equation reads:
\begin{equation}
	\label{eq3.1b}
	-\frac{1}{2}\frac{d^{2}}{dx^{2}}\psi_{{}_+}(x)+ \lambda x\psi_{{}_+}(x)=E\psi_{{}_+}(x)\;.
\end{equation}
whose solution is given by
\begin{equation}
	\label{eq3.5}
	\psi_{{}_+}(x)=N_{{}_+}\hbox{Ai} \left(\sqrt[3]{2k}\left(x-\frac{E}{\lambda}\right)\right)\;,
\end{equation}
where $N_{{}_+}$ is an appropiate normalization constant. On the other hand, if $x<0$ the Schr\"odinger equation reads:
\begin{equation}
	\label{eq3.1c}
	-\frac{1}{2}\frac{d^{2}}{dx^{2}}\psi_{{}_-}(x)- \lambda x\psi_{{}_-}(x)=E\psi_{{}_-}(x)\;.
\end{equation}
and its solution can be written as,
\begin{equation}
	\label{eq3.5b}
	\psi_{{}_-}(x)=N_{{}_-}\hbox{Ai} \left(\sqrt[3]{2k}\left(-x-\frac{E}{\lambda}\right)\right)\;,
\end{equation}
where $N_{{}_-}$ is another appropiate normalization constant.\\

From wave functions (\ref{eq3.5}) and (\ref{eq3.5b}), it is easy to see that if $\psi_{{}_+}(x)=\psi_{{}_-}(-x)$, then $N_{{}_+}=N_{{}_-}$, and we have the even wave solutions. Similarly, if  $\psi_{{}_+}(x)=-\psi_{{}_-}(-x)$, then  $N_{{}_+}=-N_{{}_-}$ and we have odd wave solutions. In order to solve equation (\ref{eq3.1}), we must connect the two wave functions (\ref{eq3.5}) and (\ref{eq3.5b}) along with their derivates at $x=0$. From $\psi_{{}_-}(0)=\psi_{{}_+}(0)$, and $\psi'_{{}_-}(0)=\psi'_{{}_+}(0)$, we write
\begin{eqnarray} 
	(N_{{}_-}-N_{{}_+})\hbox{Ai}\left(-\sqrt[3]{\frac{2}{\lambda^2}}E\right)&=&0\;,\label{eq3.5d}\\
	(N_{{}_-}+N_{{}_+})\hbox{Ai}'\left(-\sqrt[3]{\frac{2}{\lambda^2}}E\right)&=&0\;,\label{eq3.5e}
\end{eqnarray}
respectively. If $N_{{}_-}=N_{{}_+}$, from equation (\ref{eq3.5e}) we define the energy of the even states as,
\begin{equation} \label{eq3.5f}
	E_{2n}=-\sqrt[3]{\frac{\lambda^2}{2}}a'_{n+1}\;,
\end{equation}
where $a'_n$ are the $n$-th zeros of the Airy function derivate \cite{abramowitz}. Then the even states will be,
\begin{equation} \label{eq3.5g}
	\psi_{{}_{2n}}(x)=N_{{}_{2n}}\hbox{Ai}\left(\sqrt[3]{2\lambda}\left(\vert x\vert-\frac{E_{2n}}{\lambda}\right)\right)\;.
\end{equation}
Similarly, if $N_{{}_-}\neq N_{{}_+}$, from equation (\ref{eq3.5d}) we define the energy of the odd states as,
\begin{equation} \label{eq3.5h}
	E_{2n+1}=-\sqrt[3]{\frac{\lambda^2}{2}}a_{n+1}\;,
\end{equation}
where $a_n$ are the $n$-th zeros of the of Airy function \cite{abramowitz} ($n\ge 1$), and the odd states reads,
\begin{equation} \label{eq3.5i}
	\psi_{{}_{2n+1}}(x)=N_{{}_{2n+1}} \hbox{sgn}(x)\hbox{Ai} \left(\sqrt[3]{2\lambda}\left(\vert x\vert-\frac{E_{2n+1}}{\lambda}\right)\right)\;.
\end{equation}

\begin{figure}[ht!]
	\centering
	\includegraphics[width=0.8\textwidth]{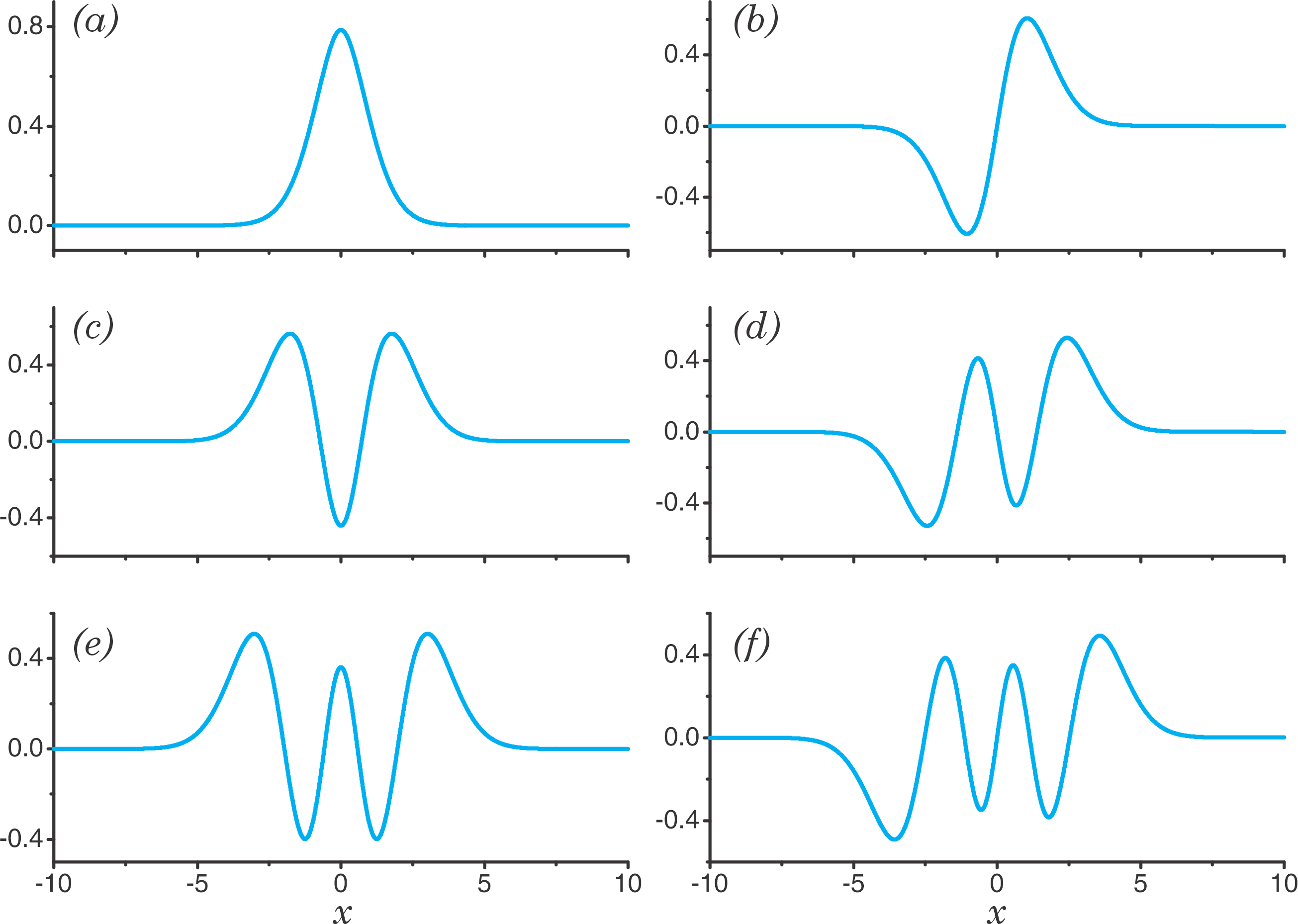} 
	\caption{The Airy eigenfunctions for six lowest energy levels.}
	\label{fig1}
\end{figure}

In figure \ref{fig1}, we show the Airy eigenfunctions for $\lambda=1$ and six lowest energy levels, (a) $E_0=$0.808616, (b) $E_1=$1.855757, (c) $E_2=$2.578096, (d) $E_3=$3.244607, (e) $E_4=$3.825715 and (f) $E_5=$4.381671, obtained from the equations (\ref{eq3.5f}) and (\ref{eq3.5h}). We have calculated numerically the normalization factors for each eigenfunction.\\ 

It is worth noting that Airy eigenfuctions were obtained from a sequence of the squeeze $\hat{S}(r)$ and displacement $\hat{D}(\alpha)$ operators, with appropriate choice for displaced parameter in terms of the energy equations (\ref{eq3.5f}) and (\ref{eq3.5h}). In case of odd functions, for $x<0$ the parity operator was used by means of sign function $\hbox{sgn}(x)$.\\

It may be numerically shown that these functions form and orthonormal basis with a weight function equal to one, i.e.,
\begin{equation} \label{eq3.5j}
	\int^{+\infty}_{-\infty}dx'\psi_m(x')\psi_n(x')=\delta_{mn}\;.
\end{equation}

\section{Applications} 

The time-dependent Sch\"odinger equation with a symmetrical linear potential
is,
\begin{equation} \label{eq3.100}
	-\frac{1}{2}\frac{d^{2}}{dx^{2}}\phi(x,t)+ \lambda\vert x\vert\phi(x,t)=i\frac{\partial}{\partial t}\phi(x,t)\;.
\end{equation}
As the Hamiltonian is time independent, the equation (\ref{eq3.100}) can be integrated with respect to time, and the formal solution for any arbitrary initial condition $\phi(x,0)$ at time $t=0$, is
\begin{equation} \label{eq3.90}
	\phi(x,t)=\exp\left[-it\left(\frac{1}{2}\hat{p}^2+\lambda\vert x\vert\right)\right]\phi(x,0)\;.
\end{equation}
Therefore, $\phi(x,0)$ can be expanded as
\begin{equation} \label{eq.392}
	\phi(x,0)=\sum^{\infty}_{n=0}c_n\psi_n(x)\;,
\end{equation}
where $\psi_n(x)$ are the Airy eigenfunctions and $c_n$ are the expansion coefficients. Considering the expansion (\ref{eq.392}), we have the propagated wave function
\begin{equation} \label{eq3.101}
	\phi(x,t)=\sum^{\infty}_{n=0}c_n\exp(-itE_n)\psi_n(x)\;,
\end{equation}
where $E_n$ are the corresponding eigenvalues of the orthogonal Airy eigenfunctions. The expansion coefficients are determined from the initial condition as
\begin{equation} \label{eq3.102}
	c_n=\int^{+\infty}_{-\infty}dx'\phi(x',0)\psi_n(x')\;,
\end{equation}
which we may obtain numerically. 

\subsection[revival]{Collapses and revivals}

Because of their charge neutrality and long lifetime, neutrons were promising candidates with which to observe experimentaly  gravitational quantum bound states \cite{Nesvizhevsky2002}. The dynamics of a bouncing wave packet under the influence of a gravity potential has been studied in Ref. \cite{gea_1999}, and the falling packet or neutrons do not move continuously along the vertical direction, but rather jump from one height to another, as predicted by quantum theory with the colapses and revivals \cite{PhysRevLett.44.1323}.\\

\begin{figure}[ht!]
	\centering
	\includegraphics[width=0.8\textwidth]{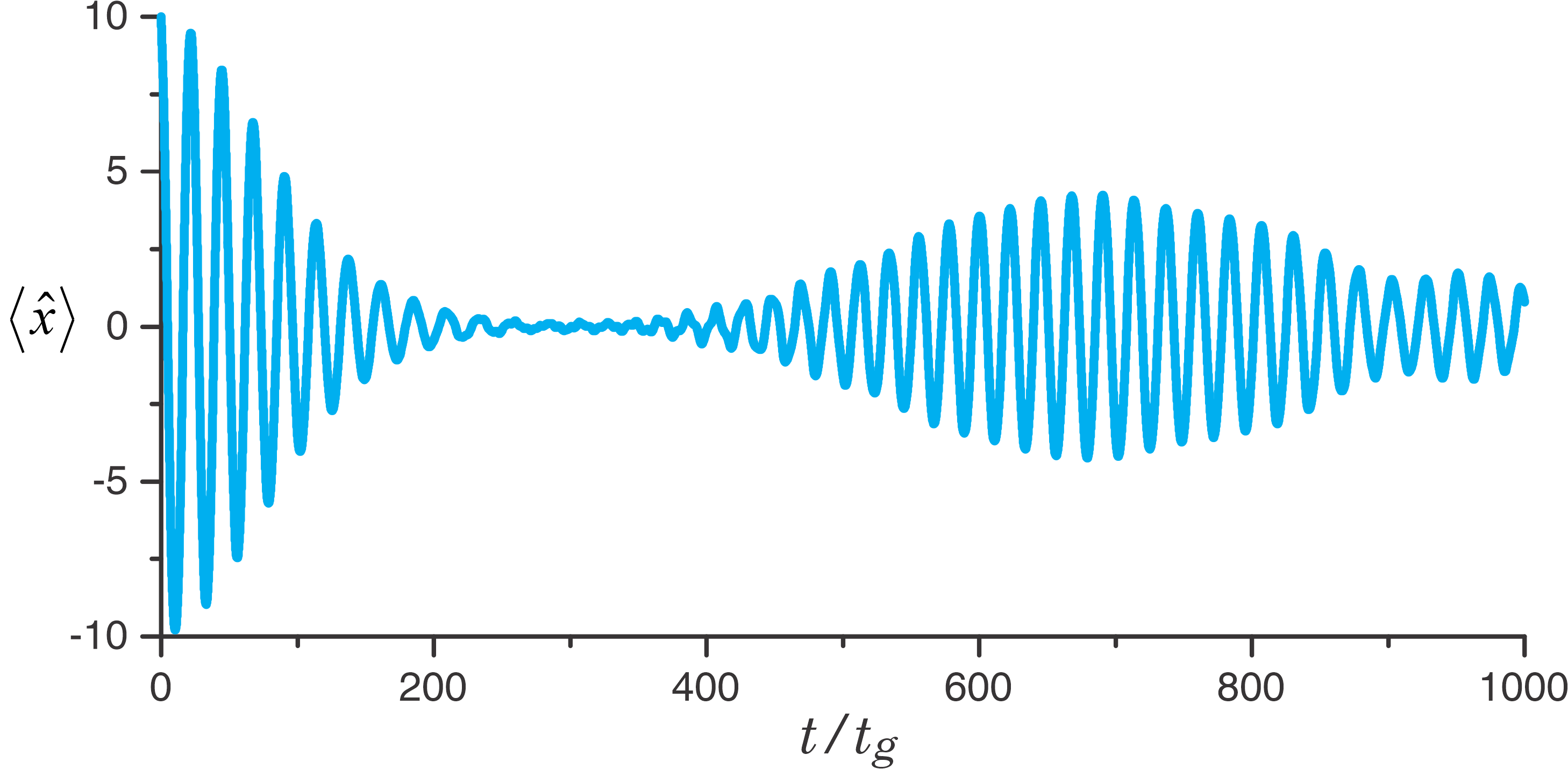} 
	\caption{Mean value of the position as a funtion of time for a wave packet defined in equation (\ref{eq3.103}).}
	\label{fig2}
\end{figure}

By employing a similar methodology as in reference \cite{gea_1999}, we consider an initial state as a Gaussian wave packet with a width $\sigma$ and localized at $x=x_0$ with zero initial momentum,
\begin{equation} \label{eq3.103}
	\phi(x,0)=\left(\frac{2}{\pi\sigma^2}\right)^{1/4}\exp\left[-\frac{(x-x_0)^2}{\sigma^2}\right]\;,
\end{equation} 
in a symmetrical linear potencial for $\lambda=1$. In order to have the trayectory of this wave packet, we obtain te mean value of coordinate $x$,
\begin{equation} \label{eq3.104}
	\langle\hat x\rangle=\int^{+\infty}_{-\infty}dx'x'\vert\phi(x',t)\vert^2\;,
\end{equation}
as is shown in figure \ref{fig2}, where $\phi(x,t)$ was obtained by using equation (\ref{eq3.101}) with $x_0=10$, $\sigma=2.0$, gravity $g=2$ and $t_g=1 /\sqrt[3]{2}$, as in reference (\cite{gea_1999}).\\

The particle has a few well defined bounces, after a while the bounces cease around the mean value of position zero. But at later times the oscillations revive an the particle begins to bounce again. This revival oscillations are purely quantum and a consequence of the discrete energy levels of this problem, obtained from the equations (\ref{eq3.5f}) and (\ref{eq3.5h}), see figure \ref{fig1}. 

\subsection[light]{Light propagation}

Laser beam propagation has been a subject of active research for several years. Quantum mechanics describes similar physics as the physical optics, providing new methods to describe and characterize the beam propagation. Paraxial wave propagation of Airy beams has been the object of several investigations. In particular, the Huygens-Fresnel integral yields the propagated wave function Airy-type wavelets \cite{Torre_2015}, defined as product of two Airy functions, i.e.,
\begin{equation} \label{eq3.105}
	E(x)=\sqrt{C}\;\hbox{Ai}\left(x+q\right)\hbox{Ai}\left(-x+q\right)\;,
\end{equation}
where $q$ is called a shift parameter and $\sqrt{C}$ is the normalization constant. It is worth mentioning that Airy patterns are very sensitive to such shift parameter.  Airy wavelets spread faster than the Gaussian beam, for this reason we will assume a symmetrical linear GRIN medium as $n^2(x)=n^2_0(1-\alpha\vert x\vert)$, in order to overrule the spreading. The one-dimensional Helmholtz equation for a GRIN medium is,
\begin{equation} \label{eq3.106}
	-\frac{\partial^2E}{\partial z^2}=\left[\frac{\partial^2}{\partial x^2}+\tilde{k}^2n^2(x)\right]E\;,
\end{equation}
where $\tilde{k}$ is the wave number, and $n(x)$ the variable refraction index. Introducing the momentun operator $\hat p$, the the Helmholtz equation is
expressed
\begin{equation} \label{eq3.107}
	\frac{\partial^2E}{\partial z^2}=-\left[\kappa^2-\left(\hat{p}^2+2\lambda\vert x\vert\right)\right]E\;,
\end{equation}
where we have defined $\kappa=\tilde{k}n_0$ and $2\lambda=\tilde{k}^2n^2_0\alpha$, whose formal solution may written as
\begin{eqnarray} 
	E(x,z)&=&\exp\left[-iz\sqrt{\kappa^2-2\left(\frac{1}{2}\hat{p}^2+\lambda\vert x\vert\right)}\;\right]E(x,0)\;,\label{eq3.108}\\
	&\approx&\exp\left(-i\kappa z\right)\exp\left[\frac{iz}{\kappa}\left(\frac{1}{2}\hat{p}^2+\lambda\vert x\vert\right)\right]E(x,0)\;,\label{eq3.109}
\end{eqnarray}
here we have developed the square root as a first order Taylor series. Comparing the propagation operator in equation (\ref{eq3.90}) with the corresponding operator in equation (\ref{eq3.109}), it is noted that $z=-\kappa t$.\\

\begin{figure}[ht!]
	\centering
	\includegraphics[width=0.6\textwidth]{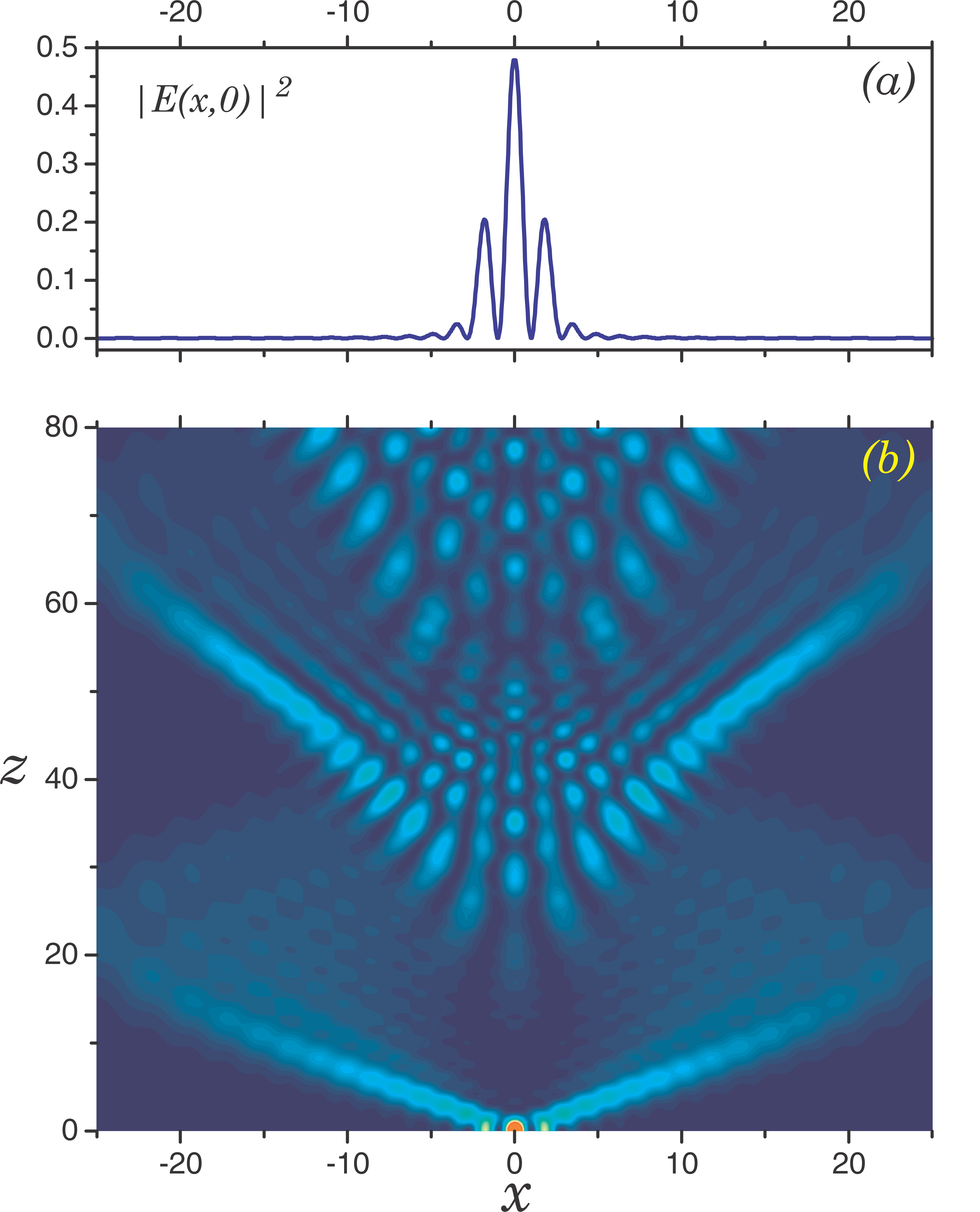} 
	\caption{(a) Initial intensity. (b) Intensity distribution of the propagation of Airy-type wavelet beam $\hbox{Ai}\left(x+q\right)\hbox{Ai}\left(-x+q\right)$ for $q=-1.472910$.}
	\label{fig3}
\end{figure}

Figure \ref{fig3}(b) depicts intensity cross-sections dynamics under symmetrical linear GRIN medium propagation of Airy-type wavelet as a function of distance $z$, which can be calculated from equation (\ref{eq3.101}) with parameters $\lambda=0.1$ and $\kappa=1.0$. The intensity of this initial field (\ref{eq3.105}) is shown in figure \ref{fig3}(a), where $q=2^{-2/3}a_1$ where $a_1$ is the first zero of the Airy function. As the propagation goes on, the initial field splits into two parts, and the intensity distribution in figure \ref{fig3}(b) is the result of interference of these two beams symmetrically reflected with respect to the plane $x=0$, and may resemble a Pearcey beam \cite{Ring:12}.

\section{Airy states in terms of position eigenstates}

Finally, in this section we will show how the Airy states may be obtained from position and momentum eigenstates. Given an arbitrary state $\langle\psi\vert$, with the completeness relation in mind, we have 
\begin{eqnarray} \label{eq4.1}
\langle\psi\vert\hbox{e}^{-ix\hat{p}}\vert\hbox{Ai}\rangle &=&   \int_{-\infty}^{+\infty}dp\;\hbox{e}^{-ixp} \langle \psi\vert p\rangle \langle p\vert\hbox{Ai}\rangle\;,\nonumber\\
&=&\int_{-\infty}^{+\infty}dp\; \langle \psi\vert\hbox{e}^{i \frac{\hat{p}^{3}}{3}}\vert p\rangle\langle p\vert x\rangle=\langle \psi\vert\hbox{e}^{i \frac{\hat{p}^{3}}{3}}\vert x\rangle\;,
\end{eqnarray}
where we have used $\exp(-ixp)=\sqrt{2\pi}\langle x\vert p\rangle$, and $\langle p\vert\hbox{Ai}\rangle=\exp(ip^3/3)/\sqrt{2\pi}$. Then
\begin{equation} \label{eq4.2}
\vert\hbox{Ai}\rangle=\exp\left(\frac{i\hat{p}^{3}}{3}\right)\exp\left(ix\hat{p}\right)\vert x\rangle\;.
\end{equation}
It is possible to check the above relation, let us consider
\begin{eqnarray} \label{eq4.3}
\langle x'\vert\hbox{Ai}\rangle &=&\langle x'\vert\exp\left(\frac{i\hat{p}^{3}}{3}\right)\exp\left(ix\hat{p}\right)\vert x\rangle\;,\nonumber\\
&=&\frac{1}{2\pi}\int_{-\infty}^{+\infty}dp\; \exp\left[i\left( \frac{p^3}{3}+x'p\right)\right]\;,
\end{eqnarray}
which recover the Airy's integral (\ref{eq001}). On the other hand, we know already that $\hat{x}\vert x\rangle=x\vert x\rangle$, from equation (\ref{eq4.2}), yield
\begin{eqnarray} \label{eq4.3a}
\hat{x}\vert x\rangle&=&\hat{x}\;\exp\left(-\frac{i\hat{p}^{3}}{3}\right)\exp\left(-ix\hat{p}\right)\vert\hbox{Ai}\rangle\;,\nonumber\\
&=&x\;\exp\left(-\frac{i\hat{p}^{3}}{3}\right)\exp\left(-ix\hat{p}\right)\vert\hbox{Ai}\rangle\;,
\end{eqnarray}
where the equation (\ref{eq0013}). The above equation allows to write position eigenstates as the application of an exponential operator to the Airy states
\begin{eqnarray} \label{eq4.3a}
\vert x\rangle=\;\exp\left(-\frac{i\hat{p}^{3}}{3}\right)\exp\left(-ix\hat{p}\right)\vert\hbox{Ai}\rangle\;,
\end{eqnarray}

Therefore Airy states may also be related to the vacuum state via an exponential operator as position eigenstates also may be written   as \cite{Soto-Eguibar2013}
\begin{equation}
    |x\rangle = \frac{e^{-x^2/2}}{\pi^{1/4}}e^{-\frac{a^{\dagger 2}}{2}+\sqrt{2}xa^{\dagger}}|0\rangle .
\end{equation}

\section{Conclusions}

In conclusion, to our best knowledge, we have obtained a new solution of the Schr\"odinger equation for a symmetrical linear potential. This solution were obtained using the squeeze $\hat{S}(r)$ and displacement $\hat{D}(\alpha)$ operators, with appropriate parameters related with the energy. This Airy ket is related to the usual coordinate eigenstate.

\end{document}